\documentclass[prc,aps,amssymb,twocolumn,showpacs]{revtex4-1} 
\usepackage{graphicx,color}
\usepackage{hyperref}
\usepackage{natbib}
\usepackage{amsmath}
\usepackage{longtable}
\usepackage{listings} 

\begin{document}
\pagebreak
\title{Shell structure from nuclear observables}
\author{I. Bentley$^{1,2}$}
\author{Y. Col\'{o}n Rodr\'{i}guez$^{1}$}
\author{S. Cunningham$^{1}$}
\author{A. Aprahamian$^{2}$}
\affiliation{$^1$Dept. of Chemistry and Physics, Saint Mary's College, Notre
Dame, IN 46556}
\affiliation{$^2$Dept. of Physics, University of Notre Dame, Notre Dame, IN
46556}

\date{\today}

\begin{abstract} 
The appearance and disappearance of shells and sub-shells are determined using a previously introduced method of structural analysis. This work extends the approach and applies it to protons, in addition to neutrons, in an attempt to provide a more complete understanding of shell structure in nuclei. Experimental observables including the mean square charge radius, as well as 
other spectroscopic and mass related quantities are analyzed for extrema. This analysis also uses 
differential observables among adjacent even-even nuclei to serve
as the derivatives for these quantities of interest. Local extrema in these quantities indicate shell 
structure and the lack of local extrema indicate missing shell closures.
The shell structure of low mass nuclei is inconsistent likely as a consequence of the single particle structure. Additionally, multiple shell features occurring in mid-shell regions are determined by combining information from two or more observables. Our results near stability complement previous observations further out.
\end{abstract}


\maketitle
\section{Introduction}

The appearance and disappearance of nuclear shells and sub-shells has been at
the forefront of recent nuclear theory and experimental efforts, see e.g.
\citep{Kanungo}-\citep{Hebeler}. Additionally, the occurrence of astrophysical
events, such as the r-process see e.g. \citep{Woosley}, depend on nuclear shell
structure to determine the location of waiting points.
Observations of shell structure near stability guide our intuition far
from stability. The goal of this work is to make use of experimental
observations of shell structure near stability to improve the predictive power further out. 

New measurements at the frontiers of the nuclear landscape indicate a scene
with evolving shells beyond the canonical magic numbers ($2,8,20,28,50,82,126$)
for neutrons ($N$) and protons ($Z$) \citep{NPNGade}. Various approaches using
nucleon-nucleon interactions \citep{Otsuka2001}, three-nucleon interactions
\citep{Otsuka3N}, tensor forces \citep{OtsukaPRL95}, super deformations
\citep{Superdeformed} and other exotic shapes, e.g. tetrahedral deformations
\citep{PRLtetrahedral}, are capable of providing explanations of the emerging
structure and new magic numbers that have been observed experimentally.

One of the most straight forward measures of a shell closure comes from the
first excited state in even-even nuclei. The first excited state is typically high in energy for a nuclide with a magic
number. Additionally, the transition probability is typically low at and near the
magic numbers. Magic numbers are also associated with enhanced stability,
therefore, corresponding nuclides have more binding energy and there is a corresponding \textquotedblleft
kink$\textquotedblright$ in the two particle separation energy. All of these
features are consequences of substantial shell gaps as discussed in Ref.
\citep{PRL99}. 

Using these metrics, new neutron shell closures such as those which occur at $N=14$, and $16$ in $^{22}$O and $^{24}$O have been observed \citep{PRL84}, \citep{SorlinPPNP} and a possible closure at $N=34$ in $^{54}$Ca \citep{GadePPNP} has been proposed.
Additionally, several shell features are known to be diminished or missing for nuclei with a canonical magic number, e.g. at $N=28$ the $^{42}$Si nucleus has a particularly low first excited state at 770 keV \citep{PRL99}, \citep{E2}. 

In the work by Cakirli, Casten, and Blaum, five 
observables and their derivatives are used to indicate neutron shell closures in regions of interest \citep{CC10}. 
The feature indicative of a shell closure for the mean-square charge radius,
$\langle r^2 \rangle$, is a flattening of values before a shell closure and a sharp
rise after. In the energies of the $2_1^+$ state a local maximum indicates a
shell closure. For the energy ratio $4^+_1$ over $2_1^+$ and $B(E2)$ values a
local minimum indicates a shell closure. 
Finite differences of adjacent data points were used to approximate the
derivative of each of these quantities which further verify the critical
points. The use of derivatives is essential when determining shell structure from two neutron
separation energies, $S_{2n}$, because they exhibit a rapid decline after crossing a
shell closure. Therefore, a minimum in the derivative of the two neutron
separation energy is the characteristic feature of a neutron shell closure.

The work discussed in this manuscript utilizes a derivatives technique similar to that in
Ref. \citep{CC10} and extends the approach. Our goal is to extend the range and scope of the shell
structure determinations and to provide new metrics for further shell structure determinations. For simplicity, we define the derivative in the same way for each observable. In this investigation, experimental data are examined for extrema to determine both proton and neutron shell closures across the
entire chart of the nuclides. We have also tested the approach with a number of
new observables. Our investigations involve determining extrema in mass related quantities: $S_{2n}$,
two proton separation
energies ($S_{2p}$), and binding energy ($B$) minus a smooth liquid drop energy ($B_{LD}$).
Additionally, the ground-state band energies of even-even nuclei from $E(2_1^+)$
up to
$E(10_1^+)$, B(E2:$2_1^+ \rightarrow 0_1^+$) and $\langle r^2 \rangle$ values
are discussed. Other quantities such as one neutron and one proton separation
energies, three point pairing gap formulas, $E(0_2^+)$, $E(2_2^+)$, $E(3_1^+)$
and various ratios of energies were also investigated, but are not included
in this manuscript due to the paucity of data and redundancy in the results.

A discussion of the methodology has been included in Sec.
\ref{sec:derivatives}.
Section \ref{sec:mass} demonstrates how mass related quantities,
specifically, two particle separation energies and binding energies can be used to determine
the location of
shell features.
Section \ref{sec:spectra} includes the results determined from using the low-lying excited states in the ground state band of even-even nuclei. 
Section \ref{sec:observations} contains a discussion of the results from
examining B(E2) and charge radii. Section \ref{sec:results} discusses
the overall results with special emphasis placed on the observations of proton shells. Overall, we find
that some shell features occur in unexpected locations and that multiple low to mid-mass nuclei
that one might expect to exhibit shell closure features simply do not. 
Finally, Sec. \ref{sec:summary} contains a
summary of the technique and the scope of its application.

\section{Determining shell features} \label{sec:derivatives}

\begin{figure*}
\begin{center}
\includegraphics[width=17.5cm]{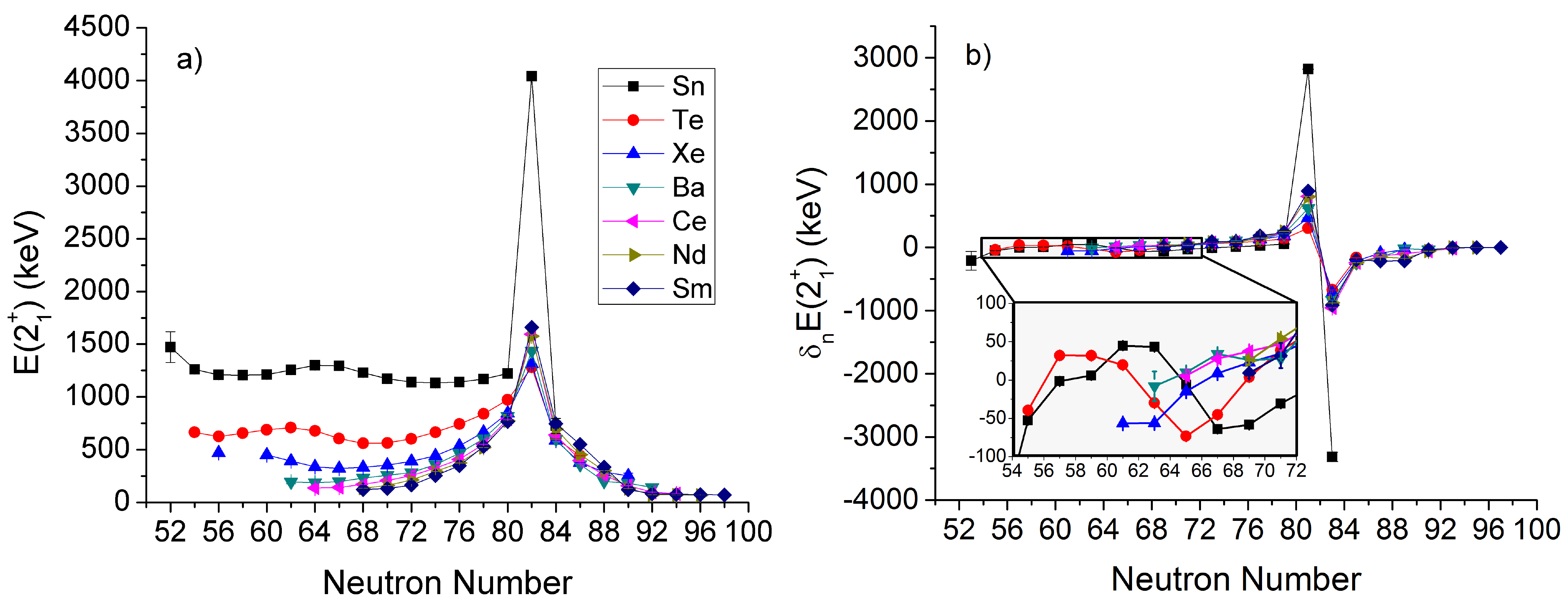}
\caption{(Color on-line) a) The first $2^+$ energy state of even-even nuclei from \citep{E2} and b) its derivative using Eq. (\ref{Eq:dn}) as a function of neutron number for five isotopes at $A \sim 130$.}
\label{fig:derivative}
\end{center}
\end{figure*}

Extrema are used to define primary and secondary signatures of shell features for various observables.
Each experimental dataset was analyzed for extrema among groups of three consecutive even-even nuclides along isotopic and isotonic chains to identify neutron and proton shell features. 
Additionally, the differences in adjacent data points were used to determine 
differential observables using the following definitions:
\begin{equation} \label{Eq:dp} 
\delta_pO(N,Z)=O(N,Z+1)-O(N,Z-1),
\end{equation} 
and 
\begin{equation} \label{Eq:dn} 
\delta_nO(N,Z)=O(N+1,Z)-O(N-1,Z),
\end{equation} 
with $O(N,Z)$ representing an experimental observable, such as $E(2_1^+)$,
$B(E2)$ and so on, for the nuclide with the corresponding
number of neutrons and protons.

The resulting $\delta_nO(N,Z)$ and $\delta_pO(N,Z)$ values are also analyzed for
extrema among three consecutive points. In the case of $S_{2n}$ and $S_{2p}$, the primary shell feature comes from the differences using Eqs. (\ref{Eq:dp}) and (\ref{Eq:dn}) and there is no secondary feature. For all other observables, extrema in the observable determines the primary shell feature signature and the derivatives before and after constitute the secondary feature.
Our procedure requires that $O(N,Z)$ be known for five consecutive nuclei so that extrema in the observable and its derivatives can be determined before and after the point of interest.

Figure \ref{fig:derivative}a) contains the energies of the
first $2^+$ state for isotopes ranging from tin to samarium in which the $N=82$ shell closure can be seen as a local maximum. The $E(2_1^+)$ values for all tin isotopes are higher than those of the other chains shown as a result of the proton shell closure at $Z=50$. 
Fig. \ref{fig:derivative}b) contains the corresponding differential
observables where the shell closure corresponds to a large positive slope before and a large negative slope afterward. In the case of doubly magic $^{132}$Sn, the derivatives at the neutron shell closure are considerably larger than the singly magic neighbors.
The shell closure at $N=82$, can be seen in both the maximum of the
energies as well as the maximum in $\delta_nE(2_1^+)$ one step before and minimum one step
afterward.

In Fig. \ref{fig:derivative}a) the primary shell signature of a maximum at $N=62$ for tellurium is far less pronounced than that of the $N=82$ closure. Additionally, for this chain the secondary feature of a drop in $\delta_nE(2_1^+)$ at $N=62$ can be seen in the inset of Fig. \ref{fig:derivative}b), but it doesn't consist of the signature maximum followed by a minimum. In cases like these the extrema in the primary feature are noted despite the lack of supporting evidence in the secondary feature. This means that some unrealistic shell features may appear in the results discussed below. Consequently, the results from multiple observables are compared to verify that each shell feature observed actually corresponds to a robustly reoccurring shell or sub-shell closure.
Furthermore, the results are inconclusive when either there are insufficient adjacent data points before or after the point of interest or if the experimental uncertainties of adjacent extrema overlap.

\section{Shells based on nuclear masses} \label{sec:mass}

\begin{figure*}
\begin{center}
\includegraphics[width=18.3cm]{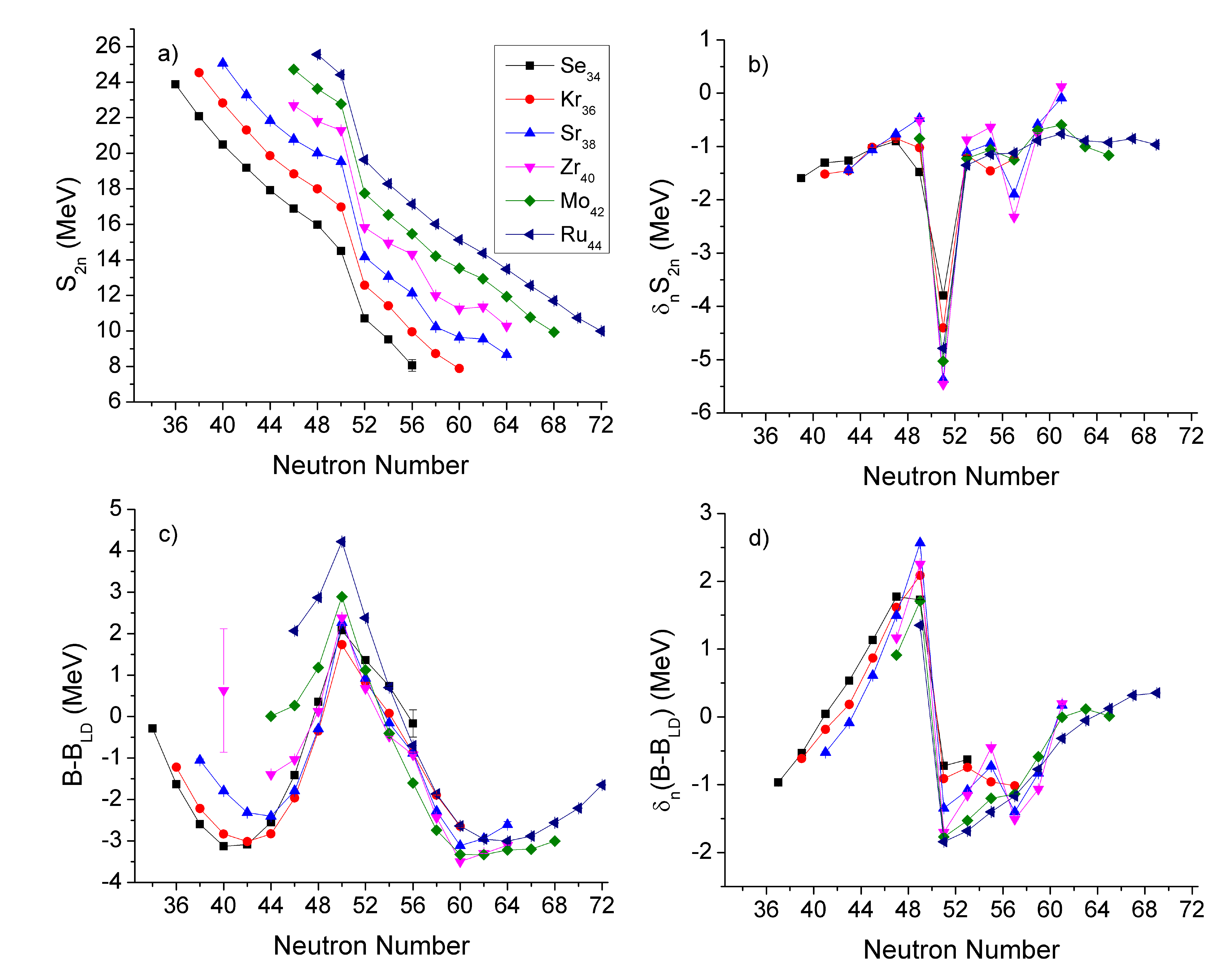}
\caption{ (Color on-line) a) Two neutron separation energy and b) its derivative from \citep{Au12} for $A \sim 90$. 
c) Binding energy minus liquid drop using Eq. (\ref{Eq:BEmLD}) and d) its derivative.}
\label{fig:SeRumass}
\end{center}
\end{figure*}

The experimentally measured binding energies, $\delta_n$$S_{2n}$ and
$\delta_p$$S_{2p}$ were taken from and calculated using data in the 2012 Atomic
Mass Evaluation (AME) \citep{Au12}. Extrapolated masses were not included in the
comparisons and the electron binding energy contribution was removed from all observables using Eq.
(A4) from Ref. \cite{Lunney03}. 

$\delta_n$$S_{2n}$ and $\delta_p$$S_{2p}$ are used to indicate the neutron and proton
shells, respectively. As a result of the definitions provided by Eqs.
(\ref{Eq:dp}) and (\ref{Eq:dn}), the minimum in the differential observable of
$S_{2p}$ and $S_{2n}$ will occur just after a shell closure. This occurs because the valence
nucleons occupy less bound orbits in a newly open shell and the separation energy drops as a consequence.

Additionally, binding energies with a liquid drop component removed can also
be used to indicate shell closures, as has been known for many years, see e.g.
\citep{MS66}. Peaks occur at magic numbers in this second comparison because magic nuclei are more tightly bound than those that are mid-shell. 
The smooth liquid drop binding energy ($B_{LD}$) that will be removed from the
experimental binding energy is of the following form:
\begin{equation}\label{Eq:BEmLD}
\begin{aligned}
B_{LD} = (a_v A+a_s A^{2/3})(1+\kappa T_Z(T_Z+1)A^{-2}) \\
+ (a_c Z(Z-1)+ \Delta)A^{-1/3}, 
\end{aligned}
\end{equation}
where $A=N+Z$ and $T_Z=(N-Z)/2$. The coefficients corresponding to a best fit
are $a_v=$ 15.79 MeV, $a_s=$ -18.12 MeV, $\kappa=$-7.18, $a_c=$-0.7147 MeV, and
$\Delta=$+5.49 MeV (for even-even nuclei). This fit corresponds to a root mean squared standard deviation of 
$\sigma=$2.65 MeV for 2353 nuclides with $N,Z>$8 in the 2012 AME \citep{Au12}.

Figure \ref{fig:SeRumass} contains mass related shell features around $N=50$ that correspond to extrema in the derivative of the two neutron separation energy and the binding energy minus liquid drop. Figures \ref{fig:SeRumass}a) and \ref{fig:SeRumass}b) illustrate the sharp decline in two neutron separation energies and the corresponding minimum in $\delta_n$$S_{2n}$ after $N=50$ and to a lesser extent after $N=56$ for strontium and zirconium.
These $N=56$ primary features are not seen in the binding energy minus liquid drop, but a secondary feature of a maximum followed by a minimum does occur in its derivative.

Overall, the primary signature results generated using separation energies and binding energies
were largely consistent with each other though more extrema were found using the derivatives of the
separation energies.
Combining the results from both of these mass related observables yields some observations of new shell features at multiple locations as can be seen in Tables \ref{tbl:AllNeutrons}-\ref{tbl:AllProtons2}.
Furthermore, the primary shell closure features are missing from both quantities for neutrons in $^{12}$Be, $^{14}$C, $^{32}$Mg, $^{34}$Si, and $^{38}$Ar and for protons in $^{18}$O and $^{42}$Ca.

A local maximum in neutrons is observed in the binding energy minus liquid drop for $N=Z$
nuclei, namely, $^{12}$C, $^{16}$O, $^{28}$Si, $^{32}$S, $^{36}$Ar, and $^{40}$Ca and for
protons in $^{28}$Si. 
Similarly, a minimum in $\delta_n$$S_{2n}$ along an isotopic chain can also be
seen for all even-even $N=Z$ nuclei from $^{12}$C to $^{44}$Ti. These results are in agreement with the findings from \citep{CC10} that $N=Z$ nuclei exhibit neutron shell features in $S_{2n}$ in the $A \sim 35$ region. Additionally,
every even-even nuclide from $^{12}$C to $^{36}$Ar was found to have a minimum in $\delta_p$$S_{2p}$ at $N=Z$.
The enhanced binding energy and drop in separation energy at $N=Z$ is likely due to enhanced proton-neutron pairing as
discussed in Refs. \citep{ZCB89}-\citep{Bentley13} and should not be considered true shell features if it doesn't persist in the other observables.

\section{Shells from the low-lying spectra of even-even nuclei}
\label{sec:spectra}

\begin{figure}
\begin{center}
\includegraphics[width=7.3cm]{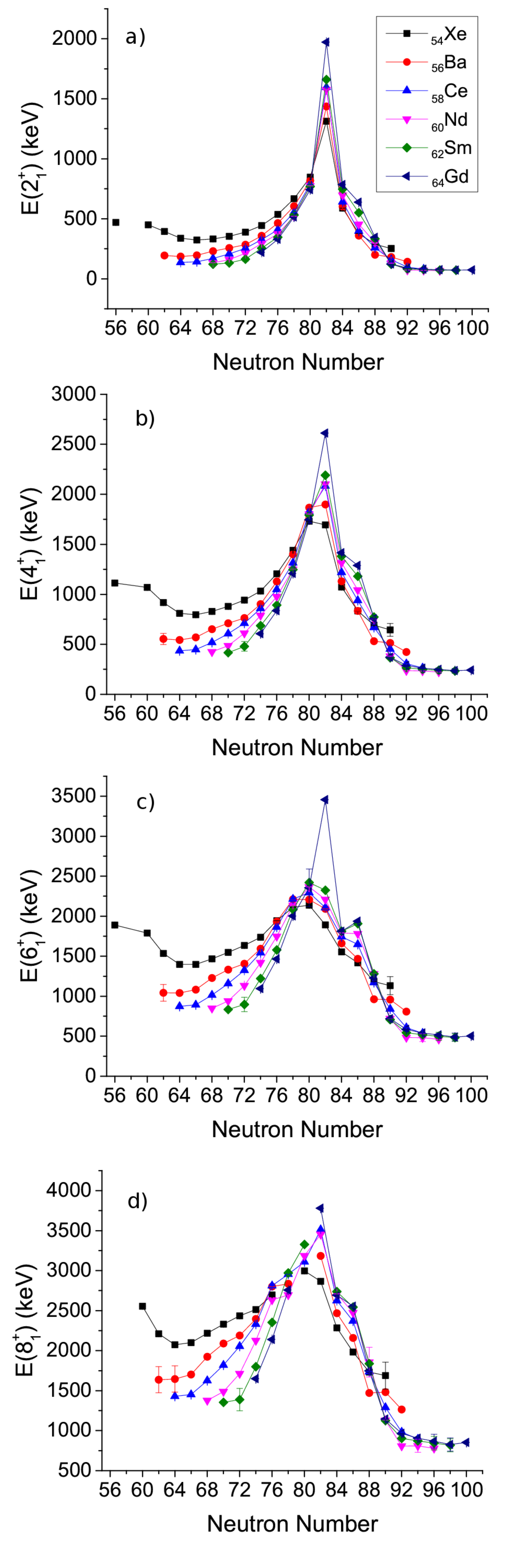}
\caption{ (Color on-line) Energies of the a) $2_1^+$, b) $4_1^+$, c) $6_1^+$, and d) $8_1^+$ states from \citep{E2} for $A \sim 140$.}
\label{fig:N82}
\end{center}
\end{figure}

Energy ratios such as $R_{4/2}=E(4_1^+)/E(2_1^+)$ can be used to investigate
shell closures. However, as opposed to using ratios, the experimental energies
for $2_1^+$, $4_1^+$, $6_1^+$, $8_1^+$ and $10_1^+$ have been analyzed
individually to provide a more complete picture of the evolving nature of shell structure in the ground-state band of even-even nuclei. 
In each case, a local maximum is the feature corresponding to a shell closure. 

Local maxima in $E(2_1^+)$ provide a list of
shell closures that are similar to those determined using mass related
quantities with the exception of the $N=Z$ nuclides which often do not contain extrema in $E(2_1^+)$. Some neutron shell closures not based on the canonical magic numbers have been found to occur in
$^{14}$C, $^{26}$Mg, $^{26}$Ne, $^{62}$Fe, $^{70}$Se, $^{68}$Zn, $^{70}$Ge, $^{68}$Ni, 
$^{94}$Sr, $^{96}$Zr, $^{110}$Cd, $^{114}$Te, $^{114}$Sn, $^{194}$Hg, and
$^{198}$Pb. 
In the case of $^{110}$Cd, for example, it is believed that shape coexistence
with a deformed 2p-4h proton excitation forms an intruder band consisting of
slightly deformed states cause shell closure-like features \citep{Garrett}.
Many of the closures listed above, such as in $^{68}$Ni at $N=40$
correspond to known, see Refs. \citep{NPNGade} and \citep{68Ni}, localized sub-shell closures based on experimental data. 
Additionally, the local maxima in $E(2_1^+)$ indicate that proton shell closures at $^{14}$C, 
$^{30}$Si, $^{34}$Si, $^{42}$Ar, $^{52}$Ti, $^{80}$Kr, $^{84}$Sr, $^{86}$Sr,
$^{146}$Gd, and $^{150}$Gd have also been found. 

Proton shell closures near $Z=20,40$ and $64$ are discussed in further detail in Sec. \ref{sec:results}, though it is worth stating that the sub-shell closure at $Z=40$ is robust, existing in five zirconium
isotopes, specifically, $^{90,92,94,96,98}$Zr. The average $2_1^+$ energy
of these five isotopes is more than three and a half times larger than the
average known energy of all other zirconium isotopes \citep{E2}. 

The majority of the shell closures indicated using $E(2_1^+)$ are also found in
$E(4_1^+)$, though the data set in the latter is smaller. Figures \ref{fig:N82}a) and \ref{fig:N82}b) demonstrate the peaks in these energies which occur at the $N=82$ shell closure. In the higher spin data,
shell closures sometimes occur at a slightly smaller proton or neutron number than before. For example,
in $E(6_1^+)$ the $N=82$ shell closure feature has in most cases moved to $N=78$ or $N=80$. Additionally, there is an overall flattening of the peak near $N=82$ as the spin increases, as can be seen in Fig. \ref{fig:N82}. 

The apparent breakdown of the $N=82$ shell at higher spin states shown in Figs. \ref{fig:N82}c) and \ref{fig:N82}d)  is another good example of where the origin of a shell feature signature is probably caused by something other than an actual shell closure. At $N=82$, higher spin states like the $6^+$ can be made by exciting nucleons into the higher spin neutron orbits, specifically the f$_{7/2}$ or h$_{9/2}$ orbitals. Below $N=82$, the $6^+$ state can't be made in the same way because only low spin neutron orbits are available. Higher orbits can be reached above the shell gap at the cost of requiring more energy. In contrast, the lower spin states $2^+$ and $4^+$ can easily be made by the available orbits \citep{Castenemail}. 
Therefore, the primary shell features for $E(6_1^+)$ and above should considered with caution and the observations of features in $E(6_1^+)$ and above have been omitted from further discussion in Sec. \ref{sec:results}

\section{Shells in other observables}\label{sec:observations}

The small deformations associated with a shell closure often occur gradually. Consequently, the B(E2:$2_1^+ \rightarrow 0_1^+$) values are typically low for several nuclei near the shell
closure and a local minimum corresponding to a magic number doesn't always stand out. 
Additionally, the data for B(E2) values found in Ref. \citep{BE2}
are somewhat sparse compared to the previously used observables.
For these reasons, only 11 shell closure features were identified and three shell closure features were determined to be missing. The only nuclides missing any evidence of an
expected closure in this observable and its derivative occur for $^{14}$C at $N=8$, $^{16}$O at $Z=8$, and $^{62}$Ni at $Z=28$. Seemingly unexpected neutron closures found are $^{68}$Ge, $^{68}$Zn, and $^{172}$Hf
at $N=36$, $N=38$ and $N=100$, respectively. 

Though there is some additional evidence for the neutron shell closures in $^{68}$Ge, $^{68}$Zn, the closure in $^{172}$Hf is not justified elsewhere. The B(E2) values used in this analysis were the most recent measurements at the time of the analysis from Refs. \citep{BE2startused}-\citep{BE2endused}.  An investigating of B(E2) values was performed for $^{172}$Hf and neighboring nuclides based on prior data from Refs.  \citep{BE2startold}-\citep{BE2endold} and newer measurements from Refs.  \citep{BE2startnew}-\citep{BE2endnew}. Subsequently, most of the measurements, including the most recent of the B(E2:$2_1^+ \rightarrow 0_1^+$) values for $^{172,174,176}$Hf indicate that there is not a substantial low point at $N=100$ \citep{BE2endnew} and therefore there is really no shell closure feature at that location.
In cases where the B(E2) value is the lone observable indicating a shell closure, the result should be considered with caution and in the case of $^{172}$Hf the shell closure simply does not exist in the newest measurements.

The mean square charge radii, $\langle r^2 \rangle$, values from Ref.
\citep{Angeli13} are also used, though this data set is even more sparse. A shell closure in $\langle r^2 \rangle$
corresponds to a local minimum and a sharp rise afterward. 
These minima are often very shallow and after accounting for the experimental
uncertainties possible peaks seen using $\delta_n\langle r^2 \rangle$ and
$\delta_p\langle r^2 \rangle$ are common place. As a result, no nuclides
conclusively indicate a proton shell feature and only four nuclides contain neutron shell
closure features using the minimum of $\langle r^2 \rangle$ itself. Those are
$^{24}$Ne and $^{26}$Mg at $N=14$, and $^{86}$Kr and $^{88}$Sr at $N=50$.
Shell closures are distinctly missing for a few high-mass nuclei including
$^{136}$Xe at $N=82$, $^{208}$Pb at $N=126$, $^{114}$Sn at $Z=50$ and $^{198,200,202}$Pb at $Z=82$, but the previously discussed evidence indicates that these shells are present. Therefore, these discrepancies from the expected shell closures may
indicate that our local extrema determination method is not well suited for
use with $\langle r^2 \rangle$ values. 

\section{Results} \label{sec:results}
\begin{figure*}
\begin{center}
\includegraphics[width=18cm]{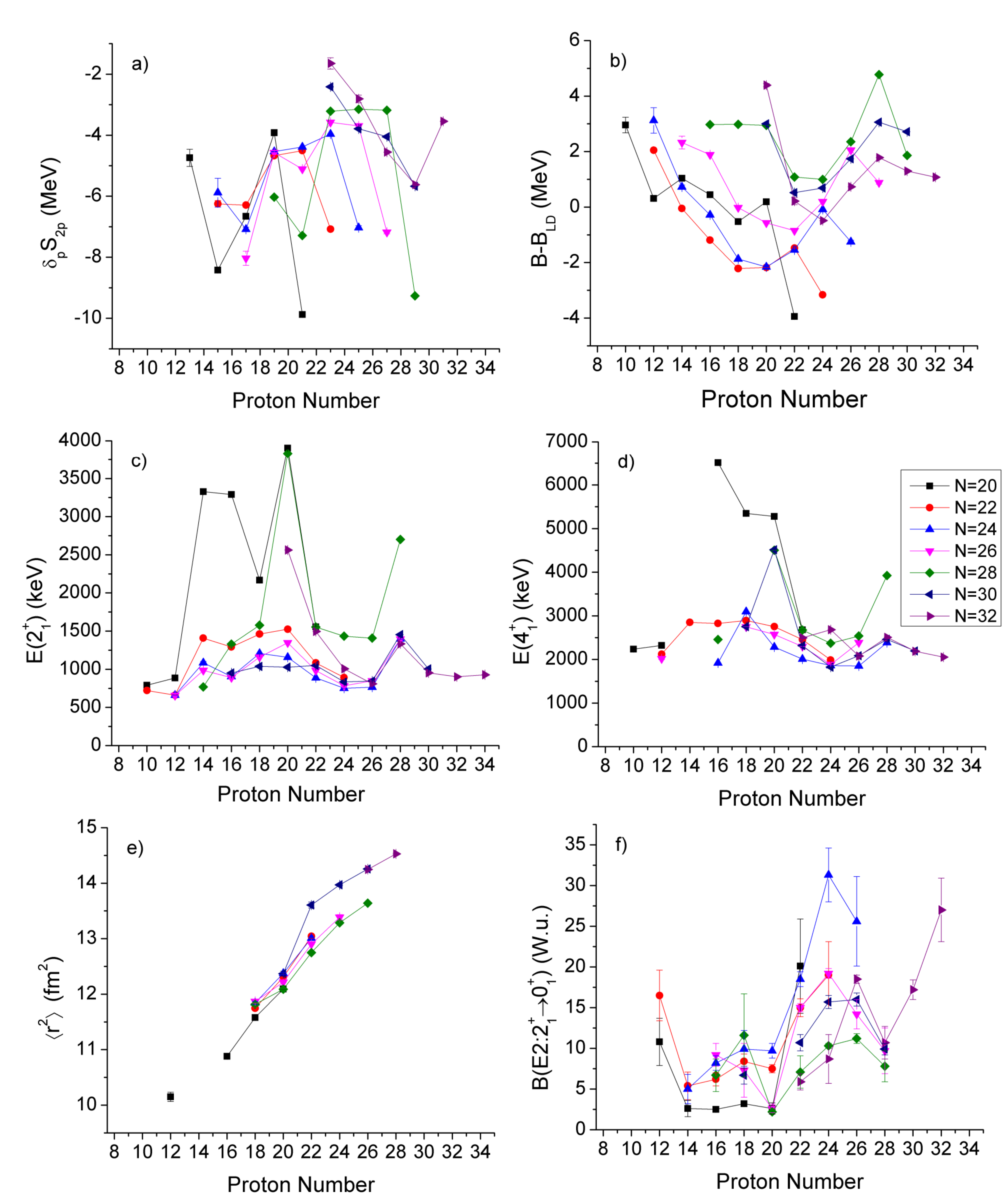}
\caption{ (Color on-line) a) Derivative in the two proton separation energy and b) binding energy minus liquid drop from \citep{Au12} for $A \sim 50$. Energies of the first excited c) $2^+$ and d) $4^+$ from \citep{E2}. e) Mean squared charge radius from \citep{Angeli13} and f) B(E2) values from \citep{BE2}. }
\label{fig:Z20}
\end{center}
\end{figure*}

\begin{figure*}
\begin{center}
\includegraphics[width=18cm]{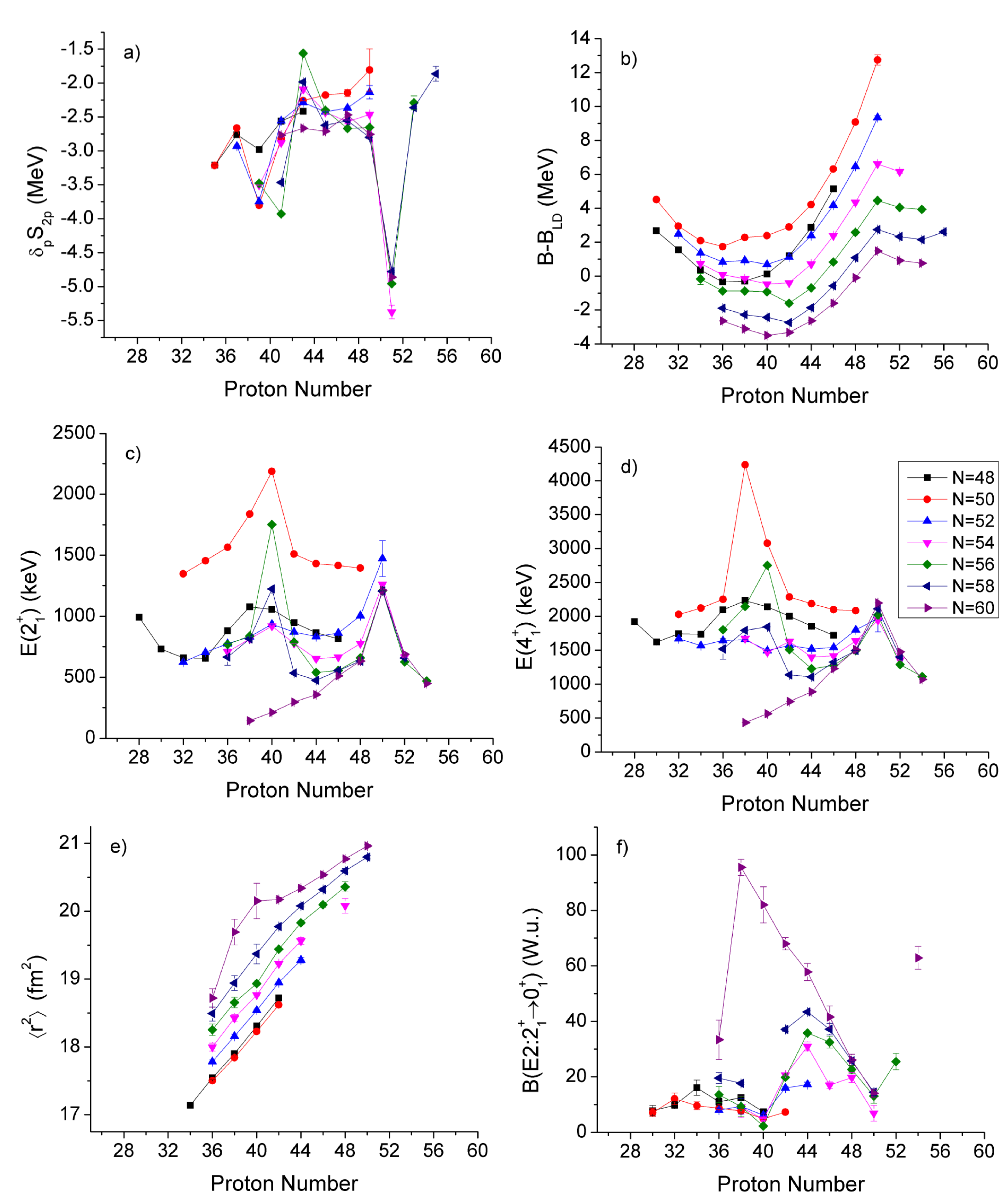}
\caption{ (Color on-line) a) Derivative in the two proton separation energy and b) binding energy minus liquid drop from \citep{Au12} for $A \sim 90$. Energies of the first excited c) $2^+$ and d) $4^+$ from \citep{E2}. e) Mean squared charge radius from \citep{Angeli13} and f) B(E2) values from \citep{BE2}. }
\label{fig:Z40}
\end{center}
\end{figure*}

\begin{figure*}
\begin{center}
\includegraphics[width=18cm]{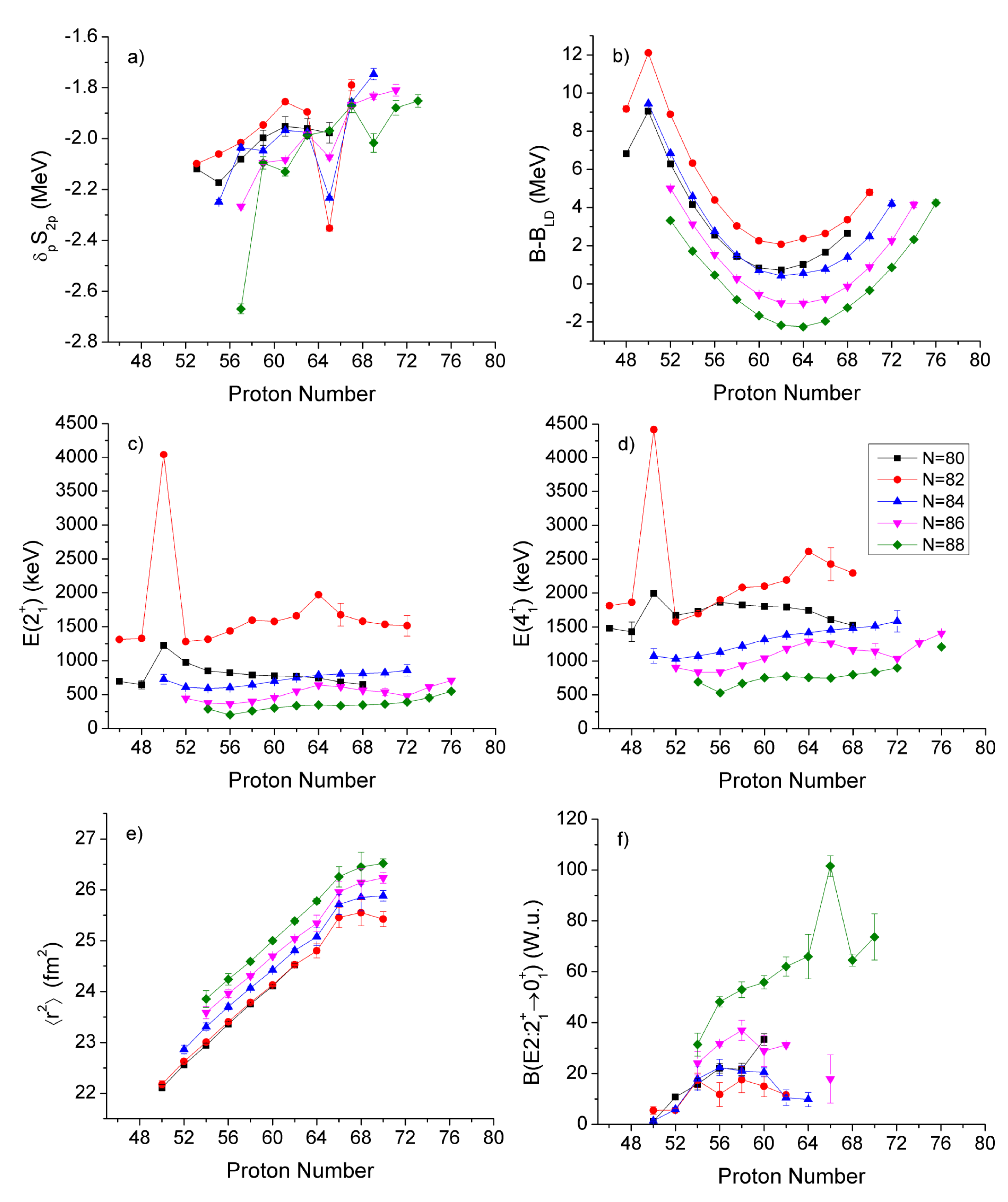}
\caption{ (Color on-line) a) Derivative in the two proton separation energy and b) binding energy minus liquid drop from \citep{Au12} for $A \sim 150$. Energies of the first excited c) $2^+$ and d) $4^+$ from \citep{E2}. e) Mean squared charge radius from \citep{Angeli13} and f) B(E2) values from \citep{BE2}. }
\label{fig:Z64}
\end{center}
\end{figure*}

Extrema in experimental observables and the corresponding differential
observables were determined by comparing groups of adjacent even-even nuclei along isotopic and
isotonic chains. The extrema indicative of neutron and proton shell structure were then used to identify nuclides of interest. 
Figs. \ref{fig:Z20}-\ref{fig:Z64} illustrate how these shell closure features occur among the six preferred observables near shell and sub-shell closures at $Z=20$, $Z=40$ and $Z=64$, respectively.

Figures \ref{fig:Z20}a) and \ref{fig:Z20}b) can be used to examine the $Z=20$ shell closure in some of the isotones shown, as well as features associated with enhanced pairing at $N=Z$. In Fig. \ref{fig:Z20}a) the rapid decrease in $\delta_p$$S_{2p}$ can be seen for the $N=Z$ which is similar to observations made along isotopic chains in Ref. \citep{CC10}. Figure  \ref{fig:Z20}b) illustrates that the closure at $Z=20$ is only clearly present in the $N=20$ chain.
Figs. \ref{fig:Z20}c) and \ref{fig:Z20}d) contain many expected and unexpected local extrema, though the scale varies greatly among them. Figure \ref{fig:Z20}c) in particular illustrates the enhancement of sub-shell features at $Z=14$ and $Z=16$ when the companion particle is closed shell for the $N=20$ chain. 
In Fig. \ref{fig:Z20}e) the sharp rise in $\langle r^2 \rangle$ values after $^{50}$Ca provides part of the required shell feature but the flattening of values before is missing. 
In Fig. \ref{fig:Z20}f) the $Z=20$ shell closure can be clearly seen in some nuclei though it often appears to be less distinct than the next shell closure at $Z=28$. Additionally, in the $N=20$ isotones the B(E2) values are consistently small from $Z=14$ through $Z=20$ indicating that these nuclides are all spherical.
The proton shell closure is distinctly missing for $^{44}$Ca across all observables. 
Overall, the $Z=20$ shell is a mixture of some features associated with shell closures and some features which are missing. 
This closure is believed to evolve as a result of tensor forces between the respective protons and neutrons \citep{SorlinPPNP}. 

Figure \ref{fig:Z40}a) illustrates some unexpected features at $Z=38$, as well as expected sub-shell features at $Z=40$ and shell features at $Z=50$. The sharp distinct drop in two proton separation energies can be seen at either $Z=38$ or $Z=40$ in the $N=48$ through $N=56$ chains depending on the isotone. Figure \ref{fig:Z40}b) only indicates the $Z=50$ closure.
It should be noted that in for both the $N=50$ and $N=56$ chains the $2^+$ energies shown in Figure \ref{fig:Z40}c) are higher at the sub-shell closure $Z=40$ than at the shell closure $Z=50$, though the shell closure at $Z=50$ is more persistent. In Figs.\ref{fig:Z40}c) and \ref{fig:Z40}d) the peak in the $N=50$ chain shifts from $Z=40$ in $E(2_1^+)$ to $Z=38$ in $E(4_1^+)$.
The sharp rise in charge radius values at $Z=36$ in Figure \ref{fig:Z40}e) for the $N=60$ chain and others, are inconclusive because of the lack of data at lower neutron numbers. Similarly, the flattening out and then increase as seen in the $N=60$ chain near $Z=40$ is inconclusive as a result of the considerable experimental uncertainties.
Figure \ref{fig:Z40}f) shows that many of the B(E2) values in the $Z=30-40$ region are small.
Figures \ref{fig:Z40}c)-\ref{fig:Z40}f) also demonstrate the consequences for various observables as the deformation decreases along the $N=60$ chain.

A distinct drop in two proton separation energies can be seen in Fig. \ref{fig:Z64}a) at $Z=64$ for $^{146,148,150}$Gd. The isotones shown in Fig. \ref{fig:Z64}b) only indicate the shell closure at $Z=50$.
Figures \ref{fig:Z64}c) and \ref{fig:Z64}d), show peaks at $Z=50$ for two of the chains. Additionally, $^{146}$Gd contains a distinct peak for both E($2^+_1$) and E($4^+_1$), while the peaks in these two quantities at $^{150}$Gd are more modest. 
Figures \ref{fig:Z64}e) includes a slight upward kink at $Z=64$ for the chains shown, though the flattening feature before was missing.
For these isotones the B(E2) data is sparse. However, the low values near $Z=64$ among the $N=82,84$ and $86$ chains, resulting from the $N=82$ shell closure, reinforce the notion of a sub-shell closure corresponding to a small deformation as can be seen in Fig. \ref{fig:Z64}f) .

In summary, the proton sub-shell closures at $Z=40$ are in agreement with
calculations by Otsuka et al., which indicate that the substantial gap between
the p$_{1/2}$ and
g$_{9/2}$ proton orbitals is caused by tensor forces \citep{Otsuka2010}. This
shell closure and another at $Z=64$ are both detected using signatures in extrema as it is indicated
by $\delta_pS_{2p}$, and
across the low-lying spectra. 
Additionally, a neutron sub-shell closure at
$N=56$ for $^{94}$Sr,$^{96}$Zr and $^{98}$Mo is similarly
indicated by $\delta_n$$S_{2n}$ and spectra. 

Interestingly, all of these more persistent sub-shell cases occur
at or near nuclides with a shell closure in the companion particle, $N=50$,
$N=82$, or the sub-shell closure at $Z=40$, respectively. 
These observations indicate that the two critical criteria needed for
the creation of a sub-shell structure are (i) a shell closure in the companion
particle and (ii) a change in spin and parity. 
Take for example, the proton sub-shell closure at $Z=64$ observed in $^{146}$Gd and $^{150}$Gd. The companion neutrons are at or near closed shells with $N=82$ and $N=86$, respectively, and the odd-proton spin-parity changed in the neighboring europium and terbium isotones from $5/2^+$ to $1/2^+$. Though the change around $Z=64$ in spin and parity is not as drastic as the more prototypical change around $Z=40$, from $1/2^-$ to $9/2^+$ for $^{88-98}$Zr, it appears to have had a sufficient effect.

One can think of the first criterion as being conducive for enhancing features because a nearby shell closure in the companion particles often results in small deformations, causing large gaps in the single particle spectra, which enhance stability and cause the ground state band to be higher in energy. A prescription based on these observations can be used to predict new sub-shell features in emerging data further from stability. But it appears that the rules for both shells and sub-shells may be more stringent further from stability, where for example, doubly magic $^{132}$Sn doesn't exhibit neutron shell quenching but neighboring nuclei do \citep{JYFLTRAP2012}.

A handful of nuclides with a magic neutron number are missing shell features across multiple observables
including $^{14}$C and $^{32}$Mg. Nucleon-nucleon
interactions may be responsible for the disappearance of shells and the emergence
of others in $^{14}$C and other low mass nuclides \citep{Otsuka2001}. For $^{32}$Mg, a two particle-two hole
configuration occurs eliminating the $N=20$ shell as discussed in Refs. \citep{Sorin13},
\citep{Janssens} and references therein. As a consequence, the deformed ground-state of this
nuclide results in a comparatively low $2^+_1$ state. 

Tables \ref{tbl:AllNeutrons}-\ref{tbl:AllProtons2} summarize all nuclei where the primary signature of a 
shell closure, i.e. a maximum or minimum, has been identified across the nine observables used. It should be noted that the features included have not been separated by their relative magnitude. Instead the table simply indicates that the extremum of interest has been identified.

Tables
\ref{tbl:MissingNeutrons}-\ref{tbl:MissingProtons} contain the list of all nuclides with canonical magic numbers that contain neither primary nor secondary shell features. Nuclides have not been included in any of the tables if a secondary feature has been found even when the primary feature is missing and they have not been included if there was insufficient data.
For example, if an extrema is indicated in the derivative but not E($2^+_1$) itself, then it will not be labeled as found. Similarly, the extrema are not labeled if the experimental uncertainties at that point and an adjacent point overlap.

Many of the new shell features are distinctly
different than the canonical shells. These features often occur in just a few
observables and often last for just a few nuclides. Occasionally, the new shells migrate to a
new location such as the $N=14$
and $N=16$ sub-shells seen in oxygen as discussed in Ref. \citep{PRC69} and
citations therein.

\begin{figure*}
\begin{center}
\includegraphics[width=15.2cm]{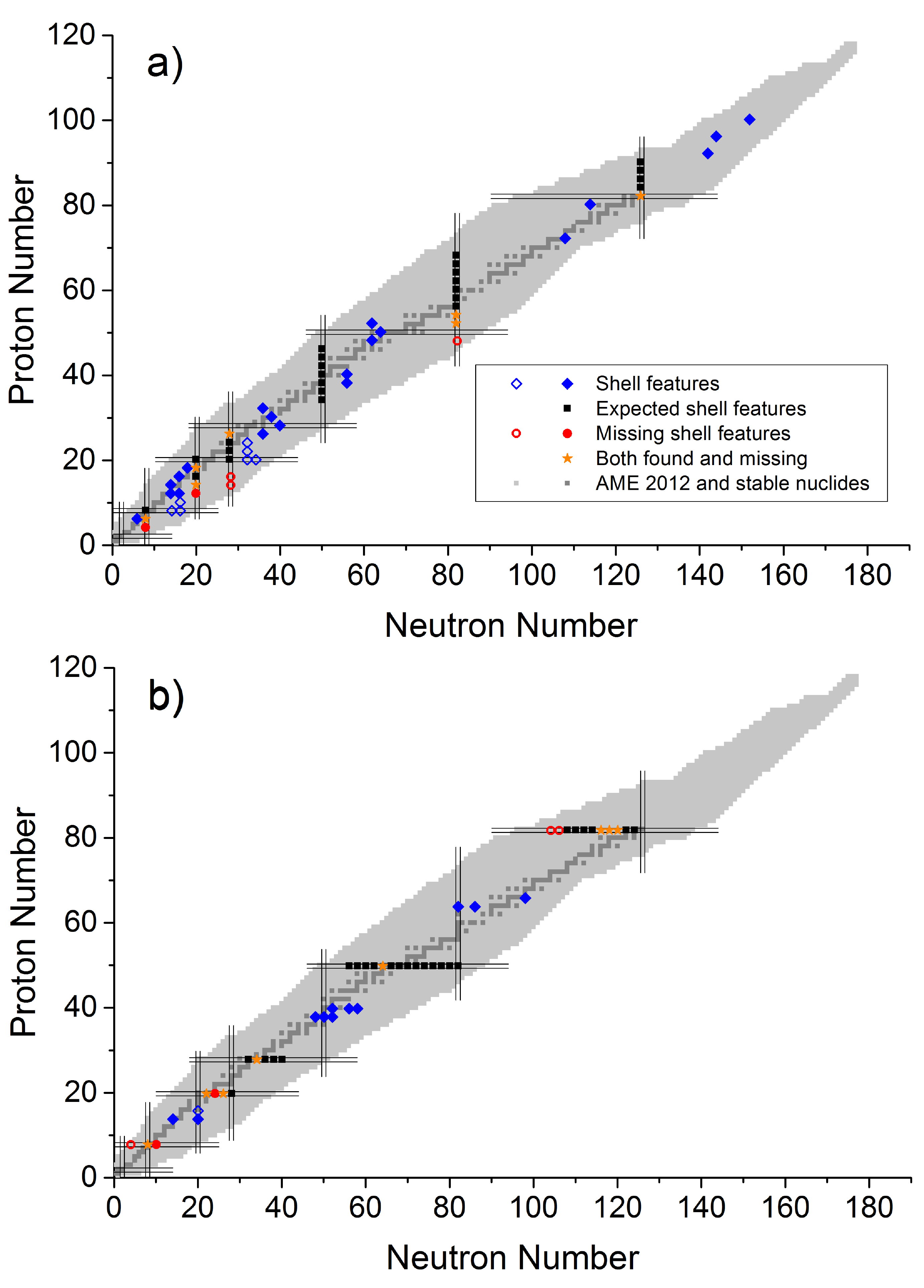}
\caption{(Color on-line) a) Neutron and b) proton shell features
from $S_{2n}$ or $S_{2p}$, $B_{Exp.}-B_{LD}$, $E(2_1^+)$, and $E(4_1^+)$. Blue diamonds indicate unexpected shell features and 
black squares indicate expected shell features found in at least two of the observables.
Red circles indicate two or more expected shell features that are missing, and orange stars indicate a combination of both
found and missing shell features. Symbols with a hollow center represent additional determinations of shell structure for $^{22}$O from \citep{N14new}, $^{24}$O from \citep{N16new}, $^{54}$Ca from \citep{Nature2013},
$^{130}$Cd from \citep{N82breakdown}, $^{12}$O from \citep{Z8breakdown}, $^{36}$S from \citep{Z16new},
$^{186-188}$Pb from \citep{Z82breakdown}, and otherwise from \citep{Kanungo}-\citep{Otsuka13}.
For reference, dark gray squares indicate stable nuclides with half-lives greater than 10$^{24}$
yr based on data from Ref. \citep{E2} and the light gray squares indicate all
nuclides included in the 2012 AME \citep{Au12}.}
\label{fig:shellsummary}
\end{center}
\end{figure*}

\begin{figure*}
\begin{center}
\includegraphics[width=17.5cm]{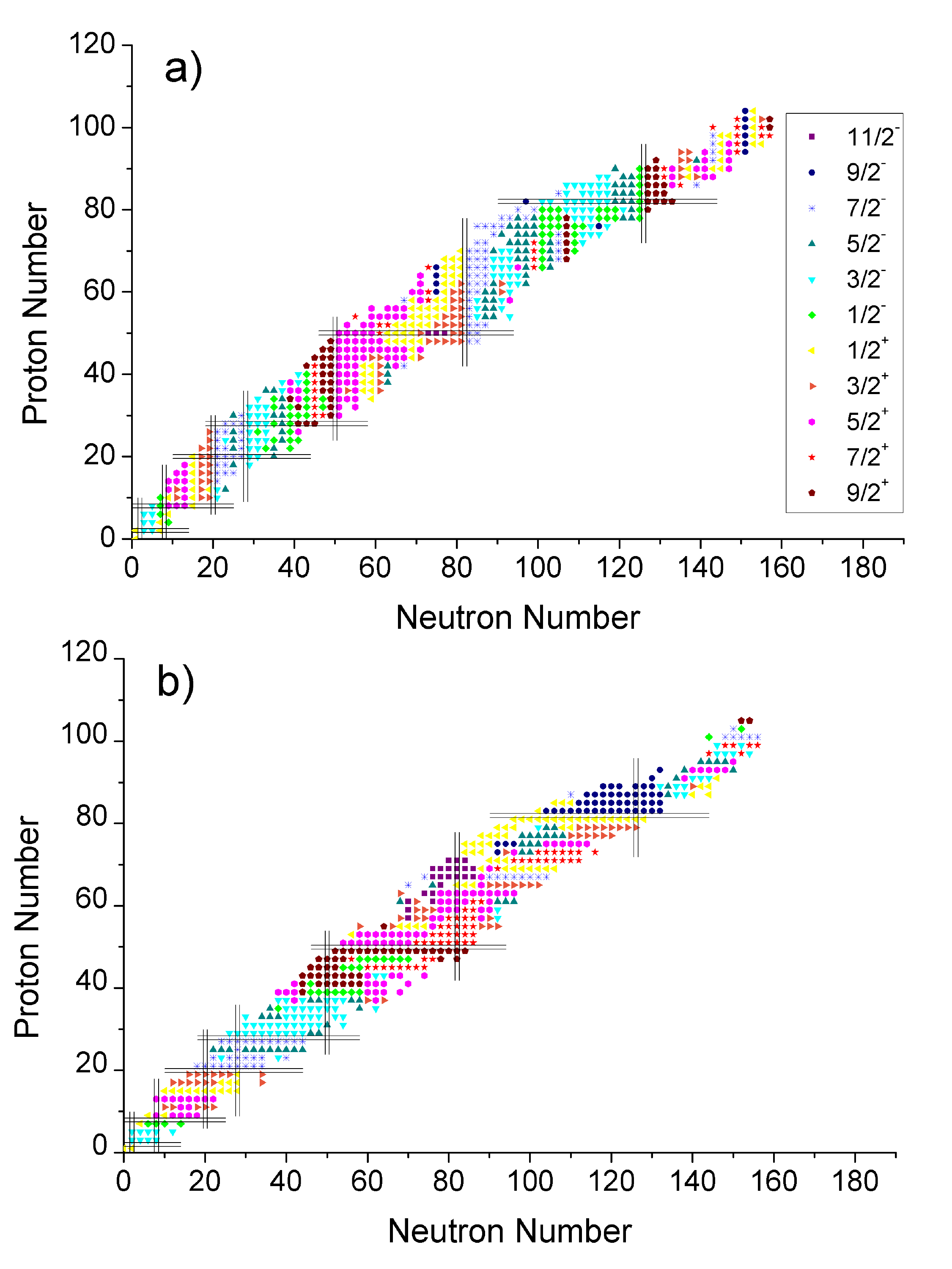}
\caption{(Color on-line) Odd-A ground-state spin and parity indicated by
color and symbol for a) odd neutron and b) odd proton nuclides with data from
\citep{E2}. }
\label{fig:OddA}
\end{center}
\end{figure*}

Figure \ref{fig:shellsummary} summarizes the shell features results
based on the combined information from all of the observables discussed in this text excluding the ground-state band
energies above $4_1^+$, B(E2) and $\langle r^2 \rangle$ values. Figure \ref{fig:shellsummary}a) includes the neutron shell features detected while Fig. \ref{fig:shellsummary}b) indicates the same for protons. The solid squares, diamonds, circles, and stars denote all nuclides with two or more shell features that are found and/or missing.

In Fig. \ref{fig:shellsummary}b) the $Z=8$ shell is less obvious when examining the amalgamated data than the $Z=20$ shell. The only observation indicating a shell closure at $Z=8$ came in $^{16}$O as a slight kink in the $S_{2p}$.
In general, missing shell features in low to mid-mass nuclei may all result from the underlying single particle structure. The expected shell closures become more consistent at and above the $N=28$ and $Z=28$ shell closures and many interesting shell features occur in mid-shell regions.

Many of the \textquotedblleft new$\textquotedblright$ features occur in at most a few adjacent nuclides. The neutron closures at $N=36, 38$, and $40$ and $N=62$, and $64$ are similar in that they occur at a 
slightly different location for the \textquotedblleft doubly magic$\textquotedblright$ nuclides than they do for the surrounding nuclides. This may be the result of the difference in tensor force interactions 
of completely closed shells and nearly closed shells.
The multiple $N=Z$ nuclides with indicated neutron shell closures below $N=20$ should be interpreted with caution
as they only occur in the mass related quantities and are likely solely a result of enhanced pairing.
Back-to-back shell closures were found at $N=14$ and $N=16$ in $^{26}$Mg and $^{28}$Mg, and at $Z=38$ and $Z=40$ in $^{90}$Sr and $^{92}$Zr, 
which both resulted from two or more shell features detected in different groups of observables.

The new and missing shell determinations from many complementary works \citep{Kanungo}-\citep{Nature2013}, \citep{N14new}-\citep{Z82breakdown} which
were often beyond the scope of our analysis, have been included Fig. \ref{fig:shellsummary}. These are denoted by open symbols. In some cases closures weren't found in our examination even though the nuclide
was within the range of nuclides examined. One such case is the $Z=16$ closure in $^{36}$S that wasn't detected because the $Z=14$ closure in $^{34}$Si was slightly more pronounced 
and was detected instead. By combining these results, shell structure for protons and neutrons has been evaluated across the chart of the nuclides.

The spin and parity in odd-A systems
can also be indicative of shell structure. 
Figure \ref{fig:OddA} has been included to allow for comparison of shell
features with the ground-state spin and parities of the adjacent odd-A
nuclides. Take for example the before mentioned $Z=40$ sub-shell
closure, which corresponds to the transition from a $1/2^-$ state to a $9/2^+$
state in the adjacent nuclides as can be seen in Fig.
\ref{fig:OddA}b).
Similarly, the transition between the $5/2^+$ state and $1/2^+$ state of the nuclides near
$^{96}$Zr correlate with the sub-shell closure at $N=56$ as can be seen in Fig.
\ref{fig:OddA}b). 

The ground-state spin and parity in odd-A nuclides do not always
provide sufficient information to allow one to consistently predict where a
shell closure will occur. For example, the exact same spin and parity transition that is seen at the
$Z=40$ shell closure also occurs for several nuclides at 
$Z=48$. In the latter case, only some of the high spin states show any indication
of a shell closure at $Z=48$ because the $Z=50$ shell closure is dominant.

\section{Summary and outlook} \label{sec:summary}

This work consists of an analysis of existing information such as $E(2^+_1)$, and $S_{2n}$, to make 
robust predictions on the appearance and disappearance of nuclear shells.
The disappearance of a shell can be produced by
particle-hole excitations within the shell model and through the restoration of
broken symmetries in mean-field approaches \citep{SorlinPPNP}.
Additionally, alternative magic numbers can be
produced in a variety of ways. For example, highly deformed nuclei and super
deformed nuclei result
in a different set of magic numbers than the canonical ones
\citep{SDps}. Although the corresponding nuclides are nominally magic, with
enhanced stability caused by considerable gaps in the single particle spectrum,
they will, by definition, not be spherical and will likely miss some spectral
features, such as a high $E(2_1^+)$ value and a low B(E2) value that are expected and looked for in this work. 
Alternative approaches such as those involving nucleon-nucleon and three nucleon interactions can explain the
emergence and disappearance of some shell features for spherical nuclei.

In principle, every shell closure should contain measurable features, but this does not mean that every feature detected, substantial or minor, corresponds with a shell closure. 
We have used a differential observable approach similar to that of Ref. \citep{CC10} to determine the location of shell closure features at a greater scale than was previously achieved.
Among the observables used to determine shell closures 
$E(2_1^+)$ and the $\delta_n$$S_{2n}$ or $\delta_p$$S_{2p}$ are among the most straightforward indicators. 
Results from the binding energy minus liquid drop supplement those from separation energies
and both detect the consequences of enhanced pairing of $N=Z$ nuclei.
The energies of higher spin states can also be used, and we show that by $6_1^+$ or higher, the peaks
begin to move away from established magic numbers, especially in the case of
$N=82$. Other observables such as
the mean square charge radii and B(E2) values can also be powerful indicators of
shell structure, but the indicative features
are often not \textquotedblleft sharp$\textquotedblright$ enough to register as an
extrema when using local comparisons.

Our local extrema determination approach is somewhat limited due to the fact that it
requires an observable to be measured in multiple adjacent nuclides. Many
results, such as missing neutron closures in $^{42}$Si \citep{Nature2005} and new neutron closures in
$^{54}$Ca \citep{Nature2013}, do not appear in Tables
\ref{tbl:AllNeutrons}-\ref{tbl:MissingProtons} as a result of the lack of data in
the neighboring nuclides away from stability. Despite of the paucity of data, we show a number of regions where new shell features are identified based on two or more experimental observations. Additionally in this work, we establish two criteria (closure in the companion particle and change in spin and parity) by which sub-shell features appear.

As experimental results continue to come in from around the world, this approach
can be repeated so that shell evolution in nuclear matter further from stability
toward the extremes of the
chart of the nuclides can be better understood. In the meantime, our approach, used in conjunction with other observations, provides the most complete picture yet of shell structure across the entire chart of nuclides.

\begin{acknowledgments}
We would like to express our gratitude to R. Casten for helpful discussions.
This work was supported by the National Science Foundation under Grants No.
PHY1419765 and No. PHY0822648.
\end{acknowledgments}

\begin{table*}
\begin{center}
\caption{Nuclides with identified signature neutron shell closure
features.\label{tbl:AllNeutrons}}
\begin{tabular}{c||c|c|c|c|c|c|c|c|c}\toprule
N&$\delta_n$$S_{2n}$&B-B$_{LD}
$&E(2$_1^+$)&E($4_1^+$)&E($6_1^+$)&E($8_1^+$)&E($10_1^+$)&$<r^2>$&B(E2)\\
\toprule
6&$^{12}$C&$^{12}$C&&&&&&&\\\hline
8&$^{16}$O&$^{16}$O&$^{14}$C&&&&&&\\\hline
10&$^{20}$Ne&&&&&&&&\\\hline
12&$^{24}$Mg&&&&&&&&\\\hline
14&$^{28}$Si&$^{28}$Si&$^{26}$Mg&$^{26}$Mg&&&&$^{24}$Ne,$^{26}$Mg&\\\hline
16&$^{28}$Mg,$^{32}$S&$^{28}$Mg,$^{32}$S&$^{26}$Ne&&&&&&\\\hline
18&$^{32}$Si,$^{36}$Ar&$^{36}$Ar&&&&&&&\\\hline
20&$^{36}$S,$^{40}$Ca&$^{40}$Ca&$^{34}$Si,$^{36}$S,&$^{36}$S&&&&&$^{36}$S,$^{38}
$Ar\\
&&&$^{38}$Ar,$^{40}$Ca&&&&&&\\\hline
22&$^{44}$Ti&&&&&&&&\\\hline
24&$^{44}$Ca&&&&&&&&\\\hline
28&$^{48}$Ca,$^{50}$Ti,&$^{50}$Ti,$^{52}$Cr,&$^{48}$Ca,$^{50}$Ti,&$^{50}$Ti,$^{
54}$Fe&&$^{50}$Ti,$^{54}$Fe&&&$^{54}$Fe\\
&$^{52}$Cr&$^{54}$Fe&$^{52}$Cr,$^{54}$Fe,&&&&&&\\
&&&$^{56}$Ni&&&&&&\\\hline
32&&&&$^{56}$Cr&$^{58}$Fe&&$^{58}$Fe&&\\\hline
34&&&&&&$^{64}$Zn&&&\\\hline
36&&&$^{62}$Fe,$^{70}$Se&$^{62}$Fe,$^{66}$Zn,&$^{66}$Zn,$^{68}$Ge&&&&$^{68}$Ge\\
&&&&$^{68}$Ge&&&&&\\\hline
38&&&$^{68}$Zn,$^{70}$Ge&$^{66}$Ni&&&&&$^{68}$Zn\\\hline
40&$^{68}$Ni&&$^{68}$Ni&&&&&&\\\hline
44&$^{76}$Ge&&&&&&&&\\\hline
46&$^{76}$Zn&&&&&&&&\\\hline
48&&&&&$^{90}$Mo,$^{92}$Ru&&&&\\\hline
50&$^{84}$Se,$^{86}$Kr,&$^{82}$Ge,$^{84}$Se,&$^{86}$Kr,$^{88}$Sr,&$^{88}$Sr,$^{
90}$Zr,&$^{90}$Zr&&$^{90}$Zr,$^{92}$Mo&$^{86}$Kr,$^{88}$Sr&\\
&$^{88}$Sr,$^{90}$Zr,&$^{86}$Kr,$^{88}$Sr,&$^{90}$Zr,$^{92}$Mo,&$^{92}$Mo,$^{94}
$Ru,&&&&&\\
&$^{92}$Mo,$^{94}$Ru&$^{90}$Zr,$^{92}$Mo,&$^{94}$Ru,$^{96}$Pd&$^{96}$Pd&&&&&\\
&&$^{94}$Ru,$^{96}$Pd&&&&&&&\\\hline
54&$^{90}$Kr,$^{102}$Cd&&&&&$^{98}$Ru&&&\\\hline
56&$^{94}$Sr,$^{96}$Zr,&&$^{94}$Sr,$^{96}$Zr&$^{94}$Sr,$^{96}$Zr&$^{96}$Zr&$^{96
}$Zr,$^{98}$Mo&$^{96}$Zr,$^{100}$Ru&&\\
&$^{98}$Mo&&&&&&&&\\\hline
58&&&&&$^{104}$Pd&$^{104}$Pd&$^{104}$Pd,$^{106}$Cd&&\\\hline
60&&&&&$^{108}$Cd,$^{112}$Te&$^{112}$Te&$^{112}$Te&&\\\hline
62&&&$^{110}$Cd,$^{114}$Te&$^{110}$Cd,$^{114}$Te&&$^{110}$Cd,$^{112}
$Sn&&&\\\hline
64&$^{106}$Mo,$^{108}$Ru,&&$^{114}$Sn&&$^{114}$Sn&&&&\\
&$^{114}$Sn&&&&&&&&\\\hline
66&&&&$^{116}$Sn&&&&&\\\hline
68&$^{112}$Ru&&&&&&$^{118}$Sn&&\\\hline
72&&&&&&$^{120}$Cd&$^{126}$Xe&&\\\hline
74&&&&&&$^{126}$Te&$^{130}$Ba&&\\\hline
76&$^{136}$Nd&&&&&&&&\\\hline
78&$^{128}$Sn&&&&$^{130}$Te,$^{134}$Ba&&&&\\\hline
80&&&&$^{132}$Te,$^{134}$Xe&$^{134}$Xe,$^{138}$Ce,&&&&\\
&&&&&$^{140}$Nd,$^{142}$Sm&&&&\\\hline
\toprule
\end{tabular}
\end{center}
\end{table*}
\begin{table*}
\begin{center}
\caption{Nuclides with identified signature neutron shell closure features
(Continued).\label{tbl:AllNeutrons2}}
\begin{tabular}{c||c|c|c|c|c|c|c|c|c}\toprule
N&$\delta_n$$S_{2n}$&B-B$_{LD}
$&E(2$_1^+$)&E($4_1^+$)&E($6_1^+$)&E($8_1^+$)&E($10_1^+$)&$<r^2>$&B(E2)\\
\toprule
82&$^{134}$Te,$^{136}$Xe,&$^{132}$Sn,$^{134}$Te,&$^{134}$Te,$^{136}$Xe,&$^{138}
$Ba,$^{140}$Ce,&$^{146}$Gd&$^{134}$Te,$^{140}$Ce,&$^{136}$Xe,$^{142}$Nd,&&$^{138
}$Ba\\
&$^{138}$Ba,$^{140}$Ce,&$^{136}$Xe,$^{138}$Ba,&$^{138}$Ba,$^{140}$Ce,&$^{142}$Nd
,$^{144}$Sm,&&$^{142}$Nd&$^{146}$Gd&&\\
&$^{142}$Nd,$^{144}$Sm,&$^{140}$Ce,$^{142}$Nd,&$^{142}$Nd,$^{144}$Sm,&$^{146}$Gd
,$^{148}$Dy&&&&&\\
&$^{146}$Gd,$^{148}$Dy,&$^{144}$Sm,$^{146}$Gd,&$^{146}$Gd,$^{148}$Dy&&&&&&\\
&$^{150}$Er&$^{148}$Dy,$^{150}$Er,&&&&&&&\\
&&$^{152}$Yb&&&&&&&\\\hline
86&$^{140}$Xe,$^{144}$Ce,&&&&&&&&\\
&$^{146}$Nd&&&&&&&&\\\hline
88&$^{144}$Ba&&&&&&&&\\\hline
90&&&&&&$^{166}$Os&&&\\\hline
92&$^{166}$W&&&&&&&&\\\hline
94&$^{156}$Sm,$^{160}$Dy,&&&&&&&&\\
&$^{162}$Er&&&&&&&&\\\hline
98&$^{164}$Dy,$^{168}$Yb&&&&&&&&\\\hline
100&$^{168}$Er,$^{182}$Pb&&&&&&&&\\\hline
102&$^{180}$Pt&&&&&&&&\\\hline
104&$^{174}$Yb,$^{176}$Hf&&&$^{180}$Os&$^{180}$Os&&&&\\\hline
106&$^{184}$Pt&&&&&&&&\\\hline
108&$^{180}$Hf,$^{182}$W,&$^{180}$Hf&&&&&$^{190}$Pb&&\\
&$^{184}$Os,$^{192}$Po&&&&&&&&\\\hline
110&&&&&&$^{190}$Hg&$^{190}$Hg&&\\\hline
112&$^{190}$Pt&&&&$^{192}$Hg&$^{196}$Po&$^{190}$Pt&&\\\hline
114&&&$^{194}$Hg&$^{194}$Hg,$^{196}$Pb&&&&&\\\hline
116&&&$^{198}$Pb&$^{200}$Po&$^{200}$Po&$^{202}$Rn&&&\\\hline
118&$^{196}$Pt&&&&$^{204}$Rn&&&&\\\hline
120&$^{200}$Hg&&&&&&&&\\\hline
126&$^{208}$Pb,$^{210}$Po,&$^{208}$Pb,$^{210}$Po,&$^{206}$Hg,$^{208}$Pb,&$^{208}
$Pb,$^{210}$Po,&$^{208}$Pb&&&&\\
&$^{212}$Rn,$^{214}$Ra,&$^{212}$Rn,$^{214}$Ra,&$^{210}$Po,$^{212}$Rn&$^{212}
$Rn&&&&&\\
&$^{216}$Th&$^{216}$Th&&&&&&&\\\hline
130&$^{214}$Po&&&&&&&&\\\hline
132&$^{218}$Rn&&&&&&&&\\\hline
134&$^{224}$Th&&&&&&&&\\\hline
138&$^{226}$Ra,$^{228}$Th,&&&&&&&&\\
&$^{230}$U&&&&&&&&\\\hline
142&$^{232}$Th,$^{234}$U&$^{234}$U,$^{236}$Pu&&&&&&&\\\hline
144&$^{240}$Cm&$^{240}$Cm&&&&&&&\\\hline
152&$^{250}$Cf,$^{252}$Fm&$^{252}$Fm&&&&&&&\\
\toprule
\end{tabular}
\end{center}
\end{table*}

\begin{table*}
\begin{center}
\caption{Nuclides with identified signature proton shell closure
features.\label{tbl:AllProtons}}
\begin{tabular}{c||c|c|c|c|c|c|c|c|c}\toprule
Z&$\delta_p$$S_{2p}$&B-B$_{LD}
$&E(2$_1^+$)&E($4_1^+$)&E($6_1^+$)&E($8_1^+$)&E($10_1^+$)&$<r^2>$&B(E2)\\
\toprule
6&$^{12}$C&&$^{14}$C&&&&&&\\\hline
8&$^{16}$O&&&&&&&&\\\hline
10&$^{20}$Ne&&&&&&&&\\\hline
12&$^{24}$Mg&&&&&&&&\\\hline
14&$^{28}$Si,$^{32}$Si,&$^{28}$Si&$^{30}$Si,$^{34}$Si&&&&&&\\
&$^{34}$Si&&&&&&&&\\\hline
16&$^{32}$S,$^{40}$S&&&&&&&&\\\hline
18&$^{36}$Ar&&$^{42}$Ar&$^{40}$Ar&&&&&\\\hline
20&$^{46}$Ca,$^{48}$Ca&&$^{42}$Ca,$^{46}$Ca,&&&&&&$^{46}$Ca,$^{48}$Ca\\
&&&$^{48}$Ca&&&&&&\\\hline
22&&&$^{52}$Ti&&&&&&\\\hline
24&&&&&$^{48}$Cr&&&&\\\hline
28&$^{60}$Ni,$^{62}$Ni,&$^{60}$Ni,$^{62}$Ni,&$^{60}$Ni,$^{62}$Ni,&$^{62}$Ni,$^{
64}$Ni,&$^{62}$Ni,$^{64}$Ni&$^{62}$Ni,$^{64}$Ni&&&$^{64}$Ni,$^{66}$Ni,\\
&$^{64}$Ni,$^{66}$Ni&$^{64}$Ni,$^{66}$Ni&$^{64}$Ni,$^{66}$Ni,&$^{66}$Ni,$^{68}
$Ni&&&&&$^{68}$Ni\\
&&&$^{68}$Ni&&&&&&\\\hline
32&$^{72}$Ge&&&&&$^{74}$Ge&&&\\\hline
34&&&&&&$^{82}$Se&&&\\\hline
36&&&$^{80}$Kr&$^{82}$Kr&$^{82}$Kr,$^{84}$Kr&&&&\\\hline
38&$^{86}$Sr,$^{88}$Sr,&&$^{84}$Sr,$^{86}$Sr&$^{86}$Sr,$^{88}$Sr,&&&&&\\
&$^{90}$Sr,$^{92}$Sr&&&$^{90}$Sr&&&&&\\\hline
40&$^{96}$Zr,$^{98}$Zr&&$^{90}$Zr,$^{92}$Zr,&$^{96}$Zr,$^{98}$Zr&&&&&$^{92}$Zr,
$^{96}$Zr\\
&&&$^{94}$Zr,$^{96}$Zr,&&&&&&\\
&&&$^{98}$Zr&&&&&&\\\hline
44&$^{96}$Ru,$^{102}$Ru,&&&&&&&&\\
&$^{104}$Ru&&&&&&&&\\\hline
46&$^{100}$Pd,$^{102}$Pd&&&&&&&&\\\hline
48&&&&&$^{104}$Cd,$^{106}$Cd,&&$^{106}$Cd&&\\
&&&&&$^{108}$Cd&&&&\\\hline
50&$^{106}$Sn,$^{108}$Sn,&$^{106}$Sn,$^{108}$Sn,&$^{106}$Sn,$^{110}$Sn,&$^{106}
$Sn,$^{110}$Sn,&$^{112}$Sn,$^{114}$Sn,&$^{110}$Sn,$^{112}$Sn,&$^{110}$Sn,$^{112}
$Sn,&&$^{112}$Sn,$^{116}$Sn,\\
&$^{110}$Sn,$^{112}$Sn,&$^{110}$Sn,$^{112}$Sn,&$^{112}$Sn,$^{114}$Sn,&$^{112}$Sn
,$^{114}$Sn,&$^{116}$Sn,$^{118}$Sn,&$^{114}$Sn,$^{118}$Sn,&$^{114}$Sn,$^{118}
$Sn&&$^{118}$Sn,$^{120}$Sn\\
&$^{114}$Sn,$^{116}$Sn,&$^{114}$Sn,$^{116}$Sn,&$^{116}$Sn,$^{118}$Sn,&$^{116}$Sn
,$^{118}$Sn,&$^{120}$Sn,$^{122}$Sn,&$^{120}$Sn,$^{132}$Sn&&&\\
&$^{118}$Sn,$^{120}$Sn,&$^{118}$Sn,$^{120}$Sn,&$^{120}$Sn,$^{122}$Sn,&$^{120}$Sn
,$^{122}$Sn,&$^{124}$Sn,$^{132}$Sn&&&&\\
&$^{122}$Sn,$^{124}$Sn,&$^{122}$Sn,$^{124}$Sn,&$^{124}$Sn,$^{126}$Sn,&$^{124}$Sn
,$^{126}$Sn,&&&&&\\
&$^{126}$Sn&$^{126}$Sn&$^{128}$Sn,$^{130}$Sn,&$^{128}$Sn
,$^{130}$Sn,&&&&&\\
&&&$^{132}$Sn&$^{132}$Sn&&&&&\\\hline
52&&&&&&&$^{122}$Te&&\\\hline
54&$^{122}$Xe,$^{124}$Xe,&&&&&&$^{126}$Xe&&\\
&$^{126}$Xe,$^{128}$Xe,&&&&&&&&\\
&$^{130}$Xe,$^{132}$Xe,&&&&&&&&\\
&$^{134}$Xe&&&&&&&&\\\hline
56&$^{144}$Ba&&&$^{136}$Ba&&&$^{130}$Ba&&\\\hline
58&&&&&$^{136}$Ce&$^{134}$Ce&&&\\\hline
60&$^{152}$Nd&&&&$^{148}$Nd&&$^{136}$Nd,$^{138}$Nd&&\\
\toprule
\end{tabular}
\end{center}
\end{table*}
\begin{table*}
\begin{center}
\caption{Nuclides with identified signature proton shell closure features
(Continued).\label{tbl:AllProtons2}}
\begin{tabular}{c||c|c|c|c|c|c|c|c|c}\toprule
Z&$\delta_p$$S_{2p}$&B-B$_{LD}
$&E(2$_1^+$)&E($4_1^+$)&E($6_1^+$)&E($8_1^+$)&E($10_1^+$)&$<r^2>$&B(E2)\\
\toprule
62&&&&$^{150}$Sm&&$^{140}$Sm,$^{146}$Sm&$^{142}$Sm,$^{146}$Sm&&\\\hline
64&$^{146}$Gd,$^{148}$Gd,&&$^{146}$Gd,$^{150}$Gd&$^{150}$Gd&$^{146}$Gd&$^{150}
$Gd&&&\\
&$^{150}$Gd&&&&&&&&\\\hline
66&$^{162}$Dy,$^{164}$Dy&$^{164}$Dy&&&$^{152}$Dy&&&&\\\hline
68&$^{156}$Er&&&&&&&&\\\hline
70&$^{172}$Yb&&&&&&&&\\\hline
72&$^{162}$Hf&&&&&&&&\\\hline
74&$^{168}$W,$^{170}$W,&&&&&&&&\\
&$^{182}$W&&&&&&&&\\\hline
76&$^{178}$Os,$^{180}$Os,&&&&&&&&\\
&$^{186}$Os,$^{188}$Os,&&&&&&&&\\
&$^{190}$Os&&&&&&&&\\\hline
80&&&&&&$^{190}$Hg,$^{192}$Hg&$^{190}$Hg&&\\\hline
82&$^{190}$Pb,$^{192}$Pb,&$^{190}$Pb,$^{192}$Pb,&$^{194}$Pb,$^{196}$Pb,&$^{194}
$Pb,$^{196}$Pb,&$^{194}$Pb,$^{196}$Pb,&$^{196}$Pb,$^{202}$Pb&$^{198}$Pb&&\\
&$^{194}$Pb,$^{196}$Pb,&$^{194}$Pb,$^{196}$Pb,&$^{198}$Pb,$^{200}$Pb,&$^{198}$Pb
,$^{200}$Pb,&$^{198}$Pb,$^{202}$Pb&&&&\\
&$^{198}$Pb,$^{200}$Pb,&$^{198}$Pb,$^{200}$Pb,&$^{202}$Pb,$^{204}$Pb,&$^{202}$Pb
,$^{204}$Pb,&&&&&\\
&$^{202}$Pb,$^{204}$Pb,&$^{202}$Pb,$^{204}$Pb,&$^{206}$Pb,$^{208}$Pb&$^{206}
$Pb&&&&&\\
&$^{206}$Pb&$^{206}$Pb&&&&&&&\\\hline
86&$^{216}$Rn&&&&&&&&\\\hline
88&$^{220}$Ra,$^{222}$Ra,&&&&&&&&\\
&$^{224}$Ra,$^{226}$Ra&&&&&&&&\\\hline
92&$^{230}$U,$^{232}$U,&&&&&&&&\\
&$^{234}$U,$^{236}$U&&&&&&&&\\\hline
98&$^{248}$Cf&&&&&&&&\\\hline
100&$^{252}$Fm,$^{254}$Fm&&&&&&&&\\
\toprule
\end{tabular}
\end{center}
\end{table*}

\begin{table*}
\begin{center}
\caption{Nuclides in which experimental data shows no indication of a neutron
shell feature.\label{tbl:MissingNeutrons}}
\begin{tabular}{c||c|c|c|c|c|c|c|c|c}\toprule
N&$\delta_n$$S_{2n}$&B-B$_{LD}
$&E(2$_1^+$)&E($4_1^+$)&E($6_1^+$)&E($8_1^+$)&E($10_1^+$)&$<r^2>$&B(E2)\\
\toprule
8&$^{12}$Be,$^{14}$C&$^{12}$Be,$^{14}$C&&&&&&&$^{14}$C\\\hline
20&$^{32}$Mg,$^{34}$Si,&$^{32}$Mg,$^{34}$Si,&&&&&&&\\
&$^{38}$Ar&$^{38}$Ar&&&&&&&\\\hline
28&$^{54}$Fe&&&&$^{50}$Ti,$^{54}$Fe&&$^{54}$Fe&&\\\hline
50&&&&&$^{92}$Mo,$^{94}$Ru&$^{92}$Mo,$^{94}$Ru&&&\\\hline
82&&&&$^{134}$Te,$^{136}$Xe&$^{134}$Te,$^{136}$Xe,&&&$^{136}$Xe&\\
&&&&&$^{138}$Ba,$^{140}$Ce,&&&&\\
&&&&&$^{142}$Nd,$^{144}$Sm&&&&\\\hline
126&&&&&$^{212}$Rn&&&$^{208}$Pb&\\
\toprule
\end{tabular}
\end{center}
\end{table*}

\begin{table*}
\begin{center}
\caption{Nuclides in which experimental data shows no indication of a proton
shell feature.\label{tbl:MissingProtons}}
\begin{tabular}{c||c|c|c|c|c|c|c|c|c}\toprule
Z&$\delta_p$$S_{2p}$&B-B$_{LD}
$&E(2$_1^+$)&E($4_1^+$)&E($6_1^+$)&E($8_1^+$)&E($10_1^+$)&$<r^2>$&B(E2)\\
\toprule
8&$^{18}$O&$^{16}$O,$^{18}$O,&&&&&&&$^{16}$O\\
&&$^{20}$O&&&&&&&\\\hline
20&$^{42}$Ca&$^{42}$Ca,$^{44}$Ca,&$^{44}$Ca,$^{50}$Ca&$^{42}$Ca,$^{44
}$Ca&$^{42}$Ca&&&&\\
&&$^{46}$Ca&&&&&&&\\\hline
28&&&&&&&&&$^{62}$Ni\\\hline
50&&&&&$^{106}$Sn,$^{110}$Sn&$^{122}$Sn,$^{124}$Sn&$^{120}$Sn&$^{114}
$Sn&\\\hline
82&&&&&&$^{194}$Pb&$^{194}$Pb&$^{198}$Pb,$^{200}$Pb,&\\
&&&&&&&&$^{202}$Pb&\\
\toprule
\end{tabular}
\end{center}
\end{table*}


\begin{thebibliography}{99}


\bibitem{Kanungo} R. Kanungo, {\it Phys. Scr.} {\bf T152}, 014002 (2013).
\bibitem{Sorin13} O. Sorlin and M.G. Porquet, {\it Phys. Scr.} {\bf T152}, 014003 (2013).
\bibitem{Janssens} R.V.F. Janssens, {\it Phys. Scr.} {\bf T152}, 014005 (2013). 
\bibitem{Otsuka13} T. Otsuka, {\it Phys. Scr.} {\bf T152}, 014007 (2013). 
\bibitem{Nature2013} D. Steppenbeck, et al., {\it Nature} {\bf 502}, pp. 207-210 (2013). 
\bibitem{PRL114} M. Rosenbusch, et al., {\it Phys. Rev. Lett.} {\bf 114}, 202501 (2015).
\bibitem{Hebeler} K. Hebeler, et al., {\it Ann. Rev. of Nucl. and Part. Sci.} {\bf 65}, pp. 457-484 (2015).
\bibitem{Woosley} S.E. Woosley, et al., {\it Rev. Mod. Phys.}{\bf 74}, 1015 (2002).
\bibitem{NPNGade} A. Gade, {\it Nuclear Physics News} {\bf 23}, pp. 10-16 (2013).
\bibitem{Otsuka2001} T. Otsuka, et al., {\it Phys. Rev. Lett.} {\bf 87}, 082502 (2001).
\bibitem{Otsuka3N} T. Otsuka, et al., {\it Phys. Rev. Lett.} {\bf 105}, 032501 (2010). 
\bibitem{OtsukaPRL95} T. Otsuka, et al., {\it Phys. Rev. Lett.} {\bf 95}, 232502 (2005). 

\bibitem{Superdeformed} N. Sharma, et al., {\it Phys. Rev. C } {\bf 87}, 024322 (2013).
\bibitem{PRLtetrahedral} J. Dudek, et al., {\it Phys. Rev. Lett.} {\bf 88}, 252502 (2002).

\bibitem{PRL99} B. Bastin, et al., {\it Phys. Rev. Lett.} {\bf 99}, 022503 (2007).

\bibitem{PRL84} A. Ozawa, et al., {\it Phys. Rev. Lett.} {\bf 84}, 5493 (2000).
\bibitem{SorlinPPNP} O. Sorlin and M.G. Porquet, {\it Prog. Part. Nucl. Phys.} {\bf 61}, pp. 602–673 (2008).
\bibitem{GadePPNP} A. Gade and T. Glasmacher, {\it Prog. Part. Nucl. Phys.} {\bf 60}, pp. 161-224 (2008).
\bibitem{E2} J. Tuli, National Nuclear Data Center: Evaluated Nuclear Structure Data File. http://www.nndc.bnl.gov/ensdf/, accessed June 2014.

\bibitem{CC10} R.B. Cakirli, et al., {\it Phys. Rev. C } {\bf 82}, 061306R (2010).

\bibitem{Au12} G. Audi, et al., {\it Chin. Phys. C} {\bf 36}, p. 1287 (2012).
\bibitem{Lunney03} D. Lunney, et al., {\it Rev. of Mod. Phys. } {\bf 75}, p. 1021 (2003).
\bibitem{MS66} W.D. Myers and W.J. Swiatecki, {\it Nucl. Phys. }{\bf 81}, p. 1 (1966).




\bibitem{ZCB89} J.Y. Zhang, R. F. Casten, and D. S. Brenner, {\it Phys. Lett. B} {\bf 227}, pp. 1-5 (1989).
\bibitem{VanI95} P. Van Isacker, D. Warner, and D. Brenner, {\it Phys. Rev. Lett.} {\bf 74}, 4607 (1995). 
\bibitem{Janecke02} J. J{\"a}necke, T. W. O'Donnell and V. I. Goldanskii, {\it Phys. Rev. C} {\bf 66}, 024327 (2002).
\bibitem{Chasman07} R.R. Chasman, {\it Phys. Rev. Lett.} {\bf 99}, 082501 (2007).
\bibitem{Bentley13} I. Bentley, and S. Frauendorf, {\it Phys. Rev. C} {\bf 88}, 014322 (2013).


\bibitem{Garrett} P. Garrett, et al., {\it Phys. Rev. C} {\bf 86}, 044304 (2012).
\bibitem{68Ni} C. Gu\'{e}naut, et al. {\it Phys. Rev. C}{\bf 75}, 044303 (2007).


\bibitem{Castenemail} R. Casten private communication in Oct. 2015.
\bibitem{BE2} B. Pritychenko, National Nuclear Data Center: Reduced Transition Probabilities. http://www.nndc.bnl.gov/be2/, accessed June 2014.

\bibitem{BE2startused} V. Werner, et al., {\it J. Phys. Conf. Ser.} {\bf 312}, 092062 (2011).
\bibitem{BE2used2} A. Costin, et al., {\it Phys. Rev. C} {\bf 74}, 067301 (2006).
\bibitem{BE2used3} V. Werner, et al., {\it J. Phys. Conf. Ser.} {\bf 205}, 012025 (2010).
\bibitem{BE2used4} H. Ejiri, and G.B. Hagemann, {\it Nucl. Phys.} {\bf A161}, 449 (1971).
\bibitem{BE2endused} Y. Tanaka, et al., {\it Phys. Rev. C} {\bf 30}, 350 (1984).

\bibitem{BE2startold} B. Bochev, et al., {\it Nucl. Phys.} {\bf A282}, 159 (1977).
\bibitem{BE2old2} H. Abou-Leila, {\it Ann. Phys. (Paris)} {\bf 2}, 181 (1967).
\bibitem{BE2old3} A.Charvet, et al., {\it J. Phys. (Paris)} {\bf 32}, 359 (1971).
\bibitem{BE2old4} H. Abou-Leila, N.N. Perrin, and J. Valentin, {\it Arkiv Fysik} {\bf 29}, 53 (1965).
\bibitem{BE2old5} J. Bjerregaard, et al., {\it Nucl. Phys.} {\bf 44}, 280 (1963).
\bibitem{BE2old6} R.M. Ronningen, et al., {\it Phys. Rev. C} {\bf 15}, 1671 (1977).
\bibitem{BE2old7} T. Hammer, H. Ejiri, and G.B. Hagemann, {\it Nucl. Phys.} {\bf A202}, 321 (1973).
\bibitem{BE2endold} D.B. Fossan, and B. Herskind, {\it Nucl. Phys.} {\bf 40}, 24 (1963).

\bibitem{BE2startnew} J.M. Regis, Univ. Cologne (2011).
\bibitem{BE2endnew} M. Rudigier, et al., {\it Phys. Rev. C} {\bf 91}, 044301 (2015).


\bibitem{Angeli13} I. Angeli and K.P. Marinova, {\it At. Data Nucl. Data Tables} {\bf 99}, pp. 69-95 (2013) and updated via private communication in Feb. 2014.
\bibitem{Otsuka2010} T. Otsuka, et al., {\it Phys. Rev. Lett.} {\bf 104}, 012501 (2010).
\bibitem{JYFLTRAP2012} J. Hakala, et al. {\it Phys. Rev. Lett.}{\bf 109}, 032501 (2012).





\bibitem{PRC69} M. Stanoiu et al., {\it Phys. Rev. C} {\bf 69}, 034312 (2004).

\bibitem{N14new} E. Becheva, et al., {\it Phys. Rev. Lett.} {\bf 96}, 012501 (2006).
\bibitem{N16new} K. Tshoo, et al., {\it Phys. Rev. Lett.} {\bf 109}, 022501 (2012).
\bibitem{N82breakdown} I. Dillmann, et al., {\it Phys. Rev. Lett.} {\bf 91}, 162503 (2003).
\bibitem{Z8breakdown} D. Suzuki, et al., {\it Phys. Rev. Lett.} {\bf 103}, 152503 (2009).
\bibitem{Z16new} P.D. Cottle and K.W. Kemper, {\it Phys. Rev. C} {\bf 66}, 061301 (2002).
\bibitem{Z82breakdown} J. Heese, et al., {\it Phys. Lett.} {\bf B 302}, p. 390 (1993).

\bibitem{SDps} T. Bengtsson et al., {\it Phys. Scr.} {\bf 24}, pp. 200-214 (1981).
\bibitem{Nature2005} J. Fridmann, et al., {\it Nature} {\bf 435}, pp. 922-924 (2005). 



\end{thebibliography}
\end{document}